\documentclass[9pt,conference]{IEEEtran}
\IEEEoverridecommandlockouts
\usepackage{cite}
\usepackage{amsmath,amssymb,amsfonts}
\usepackage{graphicx}
\usepackage{textcomp}
\usepackage{xcolor}
\usepackage{multirow}
\usepackage{booktabs} 
\usepackage[ruled,linesnumbered]{algorithm2e}  
\SetKwComment{Comment}{$\triangleright$\ }{}
\usepackage{tabularx}
\usepackage{enumitem}
\usepackage[symbol]{footmisc}

\def\BibTeX{{\rm B\kern-.05em{\sc i\kern-.025em b}\kern-.08em
    T\kern-.1667em\lower.7ex\hbox{E}\kern-.125emX}}
\begin{document}

\title{F-CAD: A Framework to Explore Hardware Accelerators for Codec Avatar Decoding \vspace{-4pt}}
 
\author{\Large
Xiaofan Zhang$^{1}$, Dawei Wang$^{2}$, Pierce Chuang$^{2}$, Shugao Ma$^{2}$, Deming Chen$^{1}$, Yuecheng Li$^{2}$ \\
\normalsize
\textit{$^1$University of Illinois at Urbana-Champaign, $^2$Facebook Reality Labs Research}\\

\textit{\{xiaofan3, dchen\}@illinois.edu, \{dawei.wang, pichuang, shugao, yuecheng.li\}@fb.com \vspace{-8pt}}

\vspace{-4pt}
}

\maketitle

\begin{abstract}
Creating virtual avatars with realistic rendering is one of the most essential and challenging tasks to provide highly immersive virtual reality (VR) experiences. It requires not only sophisticated deep neural network (DNN) based codec avatar decoders to ensure high visual quality and precise motion expression, but also efficient hardware accelerators to guarantee smooth real-time rendering using lightweight edge devices, like untethered VR headsets.
Existing hardware accelerators, however, fail to deliver sufficient performance and efficiency targeting such decoders which consist of multi-branch DNNs and require demanding compute and memory resources.
To address these problems, we propose an automation framework, called F-CAD (\underline{F}acebook \underline{C}odec avatar \underline{A}ccelerator \underline{D}esign), to explore and deliver optimized hardware accelerators for codec avatar decoding.
%
Novel technologies include 1) a new accelerator architecture to efficiently handle multi-branch DNNs; 
2) a multi-branch dynamic design space to enable fine-grained architecture configurations;
and 3) an efficient architecture search for picking the optimized hardware design based on both application-specific demands and hardware resource constraints. To the best of our knowledge, F-CAD is the first automation tool that supports the whole design flow of hardware acceleration of codec avatar decoders, allowing joint optimization on decoder designs in popular machine learning frameworks and corresponding customized accelerator design with cycle-accurate evaluation.
Results show that the accelerators generated by F-CAD can deliver up to 122.1 frames per second (FPS) and 91.6\% hardware efficiency when running the latest codec avatar decoder. Compared to the state-of-the-art designs, F-CAD achieves 4.0$\times$ and 2.8$\times$ higher throughput, 62.5\% and 21.2\% higher efficiency than DNNBuilder \cite{zhang2018dnnbuilder} and HybridDNN \cite{ye2020hybrid} by targeting the same hardware device. 
\end{abstract}

\section{Introduction}

\begin{figure}[t]
    \centering
    \includegraphics[width=0.48\textwidth]{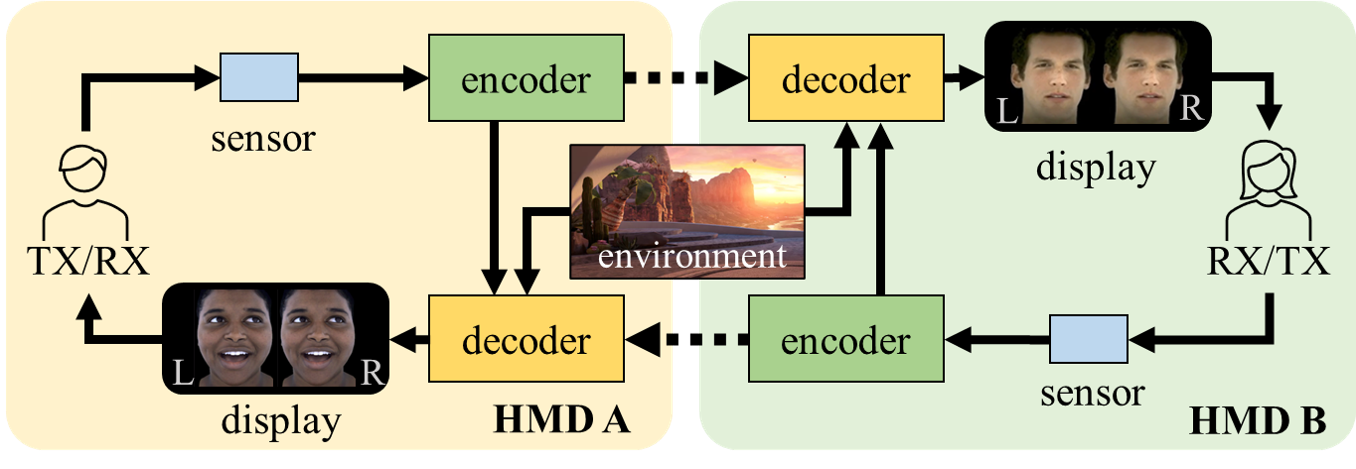}
    \vspace{-5pt}
    \caption{Social interactions with codec avatars using head-mount devices (HMDs) in VR. The TX's information is captured by the built-in sensor, compressed into a TX code by the encoder, and passed through the network. The RX then starts decoding based on the virtual environment and the state of the RX, and eventually displays TX's information on the HMD.}
    \label{fig:ca_flow}
    \vspace{-10pt}
\end{figure}

Codec avatars are photo-realistic and three-dimensional reproductions of human appearances and real-time expressions. It is one of the most impressive breakthroughs that helps people achieve VR telepresence with more effective communications with not only speaking and listening but also facial expressions and body languages \cite{lombardi2018deep,wei2019vr,chu2020expressive}. The whole system is shown in Fig. \ref{fig:ca_flow}, where all the information (e.g., a wry smile and a furrowed brow) 
of the transmitter (TX) will be encoded, transmitted, and decoded after reaching the receiver (RX) to generate the TX's codec avatar for high-fidelity social presence. Among them, the decoder is the most complex module, occupying 90\% of the calculations required by the entire system. Without effective optimizations, it can easily become the bottleneck and hinder the smooth running of VR telepresence.

Following the popularity of VR/AR headsets, social demands have been increasing and requesting real-time and high-quality codec avatar decoding. 
However, deploying codec avatar decoders on VR headsets presents significant challenges. The state-of-the-art decoders are compute- and memory-intensive. For example, the decoder we will introduce in Sec. \ref{sec:background} contains more than 13.6 GOP (Giga operations) and 7.2 million parameters, while most of the headsets can only provide limited computation, memory, and power budgets. VR applications also ask for higher refresh rate (90 or even 120 FPS) compared to non-VR applications (30 FPS) to prevent motion sickness and provide real-time response for smooth user interactions. 
It requires hardware to deliver high throughput without using large batch sizes, as the extra delay in collecting batch inputs may fail to meet the real-time requirement. In addition, emerging decoders start adopting complicated multi-branch DNNs with customized neural network layers to generate different components of the codec avatar, such as one branch for facial geometry and another for textures, and these branches may have very different requirements.
These unique challenges make codec avatar decoders difficult to be effectively handled by existing hardware accelerators. 
Evaluated by a state-of-the-art commercial SoC processor (Snapdragon 865 \cite{qualcomm865}) and two recently published DNN accelerators from academia (DNNBuilder \cite{zhang2018dnnbuilder} and HybridDNN \cite{ye2020hybrid}), we have found that all these accelerators failed to deliver satisfactory performance and efficiency required by this compelling application.
 
To address these challenges, we propose F-CAD, a new automation tool for accelerating multi-branch DNNs with complicated layer dependencies. We focus on codec avatar decoding in this paper as an important and practical use case of F-CAD to deliver the optimized hardware accelerators by meeting specific performance targets under resource budgets. To summarize, the main contributions are as follows.

\leftmargini=4mm
\begin{itemize}
    \item \vspace{-0pt} 
    This is the first work that focuses on building electronic design automation tools and providing rapid hardware accelerator design and exploration to leverage VR avatar applications for resource-constrained devices.
    \item We propose a novel elastic architecture
    to flexibly support multi-branch DNNs with complicated layer dependencies and a well-constructed architecture unit to support up to three-dimensional parallelism (3D parallelism) for high throughput and efficiency.
    \item We define a multi-branch dynamic design space to cover all possible hardware design combinations, which allows F-CAD to obtain the optimized solution with the best achievable performance.
    \item We integrate a DSE (design space exploration) engine to leverage efficient explorations within the predefined space and deliver the best accelerator by considering various customized constraints, such as available resources, maximum parallelism, maximum batch size, different branch priority, etc.
\end{itemize}

\section{Background of Codec Avatar}
\label{sec:background}

\begin{figure}[t]
    \centering
    \includegraphics[width=0.43\textwidth]{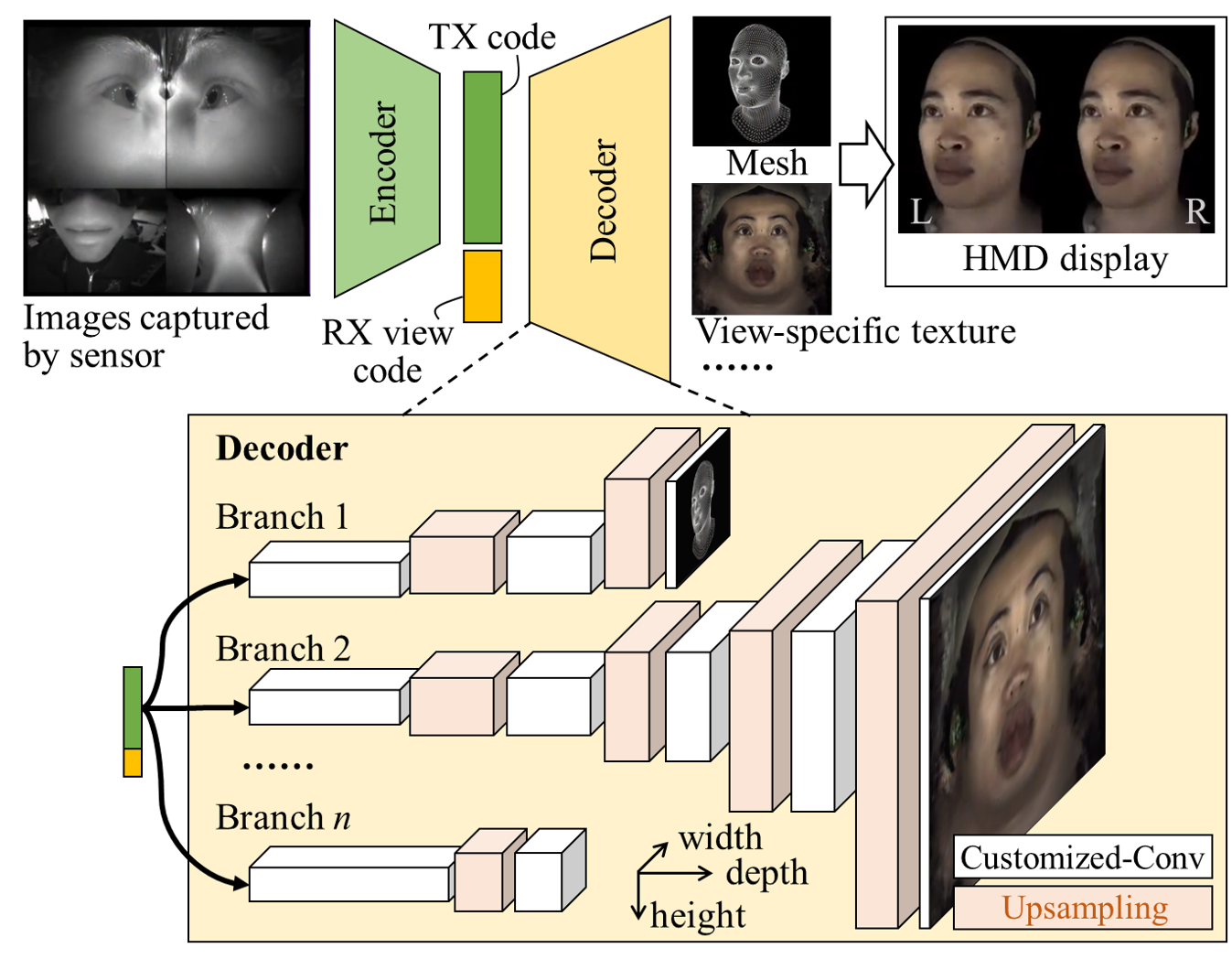}
    \vspace{-5pt}
    \caption{The pipeline of running codec avatar: from images captured by the TX to VR avatar displayed by the RX. The decoder is the most complicated part with a multi-branch DNN to generate different components of the avatar (e.g., mesh vertices in Br. 1, the view-specific texture in Br. 2, etc.).}
    \label{fig:ca_decoder}
    \vspace{-5pt}
\end{figure}

Codec avatar is formulated as a view-dependent Variational Auto-Encoder (VAE) framework \cite{lombardi2018deep,wei2019vr}. As described in Fig. \ref{fig:ca_decoder}, the encoder $E$ takes sensor-captured images $X$ as inputs and generates an \textit{l}-dimensional latent (TX) code $z$:
\begin{equation}
    z \leftarrow E(X), z \in \mathbb{R}^l
    \vspace{-5pt}
\end{equation}
where $E$ is a trained DNN transforming spatial features into a latent code. This code and the view code ($v$, indicating the RX's view direction) are processed by the decoder $D$ to generate graphic components of avatars. In Fig \ref{fig:ca_decoder}, $D$ outputs facial geometry and appearance at the first two branches:
\begin{equation}
    M, T \leftarrow D(z,v), M \in \mathbb{R}^{n\times3}, T \in \mathbb{R}^{w\times h}
    \vspace{-5pt}
\end{equation}
where $D$ is a multi-branch DNN with deconvolution-like structures for reconstructing TX's realistic VR appearances. $M$ represents the facial shape comprising $n$-vertices, and $T$ is the view-dependent RGB texture with $w\times h$ resolution. In the state-of-the-art designs, decoders are much more complicated than encoders, which contribute more than 90\% of operations of the whole VAE framework. Therefore, decoders urgently need high-performance and efficient accelerators.

In this paper, we target a state-of-the-art codec avatar decoder for facial animation. It is shown in Table \ref{tab:targeted_decoder} with three branches (Br.) for generating facial geometry (3D vertices), UV texture (a 2D surface of a 3D model following U- and V-axis), and warp field (specular effects), of which the second and third branches have a common front part. The input of Br. 1 is reshaped from a 256-dimensional latent code while the input of Br. 2 and 3 is the combination of both latent and view code to have texture appearance conditioned by different view angles.
We adopt \textbf{C}, \textbf{A}, and \textbf{U} to represent the customized Conv, activation, and up-sampling layer, respectively, while use $\times$ to indicate the number of repetitions. In total, the decoder contains 13.6 GOP and 7.2 million parameters without repeatedly counting the shared part. 
Unlike general DNNs, the decoder introduces complex data dependencies by adopting multi-branch structures and
high-definition (HD) intermediate results for high-quality VR avatar textures.
Other features come from the customized Conv, where each output pixel has its dedicated bias (aka untied bias) instead of sharing one bias across pixels within the same output channel.

\section{Accelerator Design Challenges}
\label{sec:challenges}

The unique multi-branch feature and customized layers from codec avatar decoders cause complicated dataflows and high compute and memory demands during inference, which make existing DNN accelerators ineffective. 
Challenges include the enormous (13.6 GOP) and unevenly distributed (mainly occurring in Br. 2) computations and the substantial memory footprints (with the intermediate feature map size up to $16\times1024\times1024$). This becomes even more challenging for hardware accelerators with a limited resource but aiming at real-time response with high throughput performance.

\begin{table}[t]
\footnotesize
\vspace{-5pt}
\caption{Network architecture of the targeted decoder}
\vspace{-20pt}
\label{tab:targeted_decoder}
\begin{center}
\newcommand{\tabincell}[2]{\begin{tabular}{@{}#1@{}}#2\end{tabular}}
\begin{tabular}{p{5pt}|p{32pt}|p{70pt}|c|c}
\toprule
\textbf{Br.} & \multicolumn{2}{c|}{\textbf{[Input size]}$\rightarrow$\textbf{Network}$\rightarrow$\textbf{[Output size]}} & \textbf{GOP} & \textbf{Parameters}\\ \hline
1 & \multicolumn{2}{c|}{[4,8,8]$\rightarrow$[\textbf{CAU}]$\times$5+\textbf{C}$\rightarrow$[3,256,256]} & 1.9  (10.5\%) & 1.1M (12.1\%)\\ \hline
2 & [7,8,8]$\rightarrow$  & [\textbf{CAU}]$\times$2+\textbf{C}$\rightarrow$[3,1k,1k]& 11.3  (62.4\%) & 6.1M (67.0\%) \\ \cline{3-5} \cline{1-1} 
3 & [\textbf{CAU}]$\times$5+ & \textbf{C}$\rightarrow$[2,256,256] & 4.9  (27.1\%) & 1.9M (20.9\%) \\ 
\bottomrule
\end{tabular}
\vspace{-10pt}
\end{center}
\end{table}

\begin{table}[t]
\footnotesize
\caption{Evaluations by 865 SoC (@1450MHz) \cite{qualcomm865}, DNNBuilder (@200MHz) \cite{zhang2018dnnbuilder} and HybridDNN (@200MHz) \cite{ye2020hybrid}}
\vspace{-20pt}
\label{tab:exist_rst}
\begin{center}
\newcommand{\tabincell}[2]{\begin{tabular}{@{}#1@{}}#2\end{tabular}}
\begin{tabular}{p{40pt}|c|c|c|c}
\toprule
   & \textbf{Scheme} & \textbf{Utilization} & \textbf{FPS} & \textbf{Efficiency}\\ \hline
865 (8-bit) & - & - & 35.8 & 16.9\% \\  \hline
\multirow{3}{*}{  \begin{tabular}[c]{@{}c@{}} \tabincell{l}{DNNBuilder\\(8-bit) }\end{tabular}} &  1 & DSP: 644, BRAM: 723  & 30.5& 81.6\% \\ \cline{2-5}
 & 2 & DSP: 1044, BRAM: 861 & 30.5& 50.4\%\\ \cline{2-5}
 & 3 & DSP: 1820, BRAM: 1197 & 30.5& 28.8\%\\ \hline
\multirow{2}{*}{ \begin{tabular}[c]{@{}c@{}} \tabincell{l}{HybridDNN\\(16-bit) }\end{tabular} } & 1 & DSP: 512, BRAM: 576  & 12.1  & 77.5\%  \\ \cline{2-5}
 & 2 \& 3 & DSP: 1024, BRAM: 1120 & 22.0  & 70.4\% \\
\bottomrule
\end{tabular}
\end{center}
\end{table}

\begin{figure}[t!]
\vspace{-5pt}
    \centering
    \includegraphics[width=0.42\textwidth]{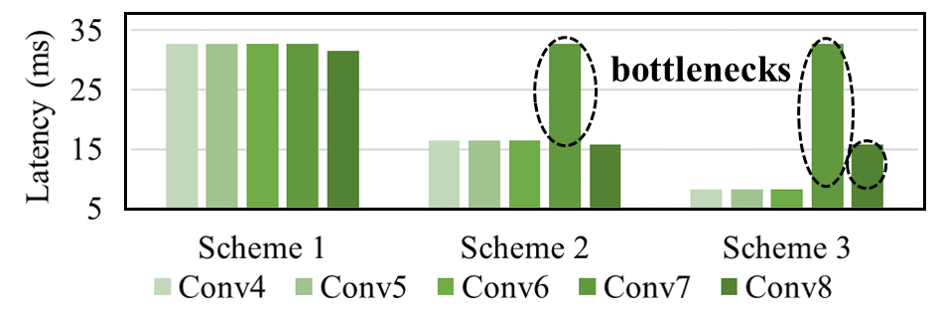}
    \vspace{-5pt}
    \caption{Latency of the last five Conv layers in Br. 2 executed by DNNBuilder when targeting FPGAs with increasing resource budgets in Scheme 1 $\sim$ 3 .}
    \label{fig:dnnbuilder_bottleneck}
    \vspace{-10pt}
\end{figure}

We select three existing accelerators from industry (Snapdragon 865 SoC \cite{qualcomm865}) and academia (DNNBuilder \cite{zhang2018dnnbuilder}, HybridDNN \cite{ye2020hybrid})
to accelerate codec avatar decoding.
For 865 SoC, we run the targeted decoder shown in Table \ref{tab:targeted_decoder}. Since DNNBuilder and HybridDNN have not supported the customized Conv, we create a mimic decoder by replacing the customized Conv with the conventional one while keeping the rest of network structure unchanged. 
The mimic decoder has a highly similar structure but 3.7\% less computations, which can still provide insights to identify bottlenecks of the existing accelerator designs.
During evaluation, we use two performance indicators including FPS (to indicate the throughput) and efficiency (the ratio between actual and theoretical peak throughput as Eq. \ref{eq:dsp_eff} to evaluate whether an accelerator works efficiently). 
$\beta$ represents the number of operations handled by one multiplier in one clock cycle. For example, $\beta=2$ for 16-bit operands in FPGA where multipliers are implemented by DSPs.

\begin{equation}
\small
\label{eq:dsp_eff}
\vspace{-2pt}
    EFFI = \frac{GOP \ per \ Second}{\beta\times \# \ of \ Multiplier\times FREQ}
\end{equation}

As shown in Table \ref{tab:exist_rst}, the 865 SoC only delivers 35.8 FPS even though it integrates an AI engine for DNN workloads. 
The major bottleneck is its limited cache size, which causes frequent data transfers and severely restricts performance. So its overall efficiency barely reaches 16.9\%. 
We then target three FPGAs (Xilinx Z7045, ZU17EG, and ZU9CG corresponding to Scheme 1, 2, and 3) with increasing resource budgets and let DNNBuilder and HybridDNN generate accelerators with unfolded and folded accelerator structures for running the mimic decoder, respectively. 
%
DNNBuilder achieves slightly lower throughput (30.5 FPS) and much higher efficiency (81.6\%) using the Z7045 FPGA in Scheme 1 compared to the 865 SoC. Its unfolded structure allows dedicated DNN layer acceleration for higher design specificity. Still, it fails to scale using more resources in Scheme 2 and 3 and suffers deteriorating efficiency. By analyzing the last five Conv layers in Br. 2, there are layers (circled in Fig. \ref{fig:dnnbuilder_bottleneck}) that can not be further accelerated due to the lack of parallelism. Regarding a Conv-like layer with $InCh$ input and $OutCh$ output channels, DNNBuilder only provides two-level parallelism with the maximum parallel factor $pf=InCh\times OutCh$. A layer with limited channel number 
(e.g., Conv7 with 16 input and output channels) 
is likely to become a computation bottleneck once DNNBuilder reaches the maximum parallel factor (e.g., $16\times16=256$) and stops scaling.
For HybridDNN, we adopt a 16-bit mimic decoder as 8-bit model is not supported. The scalability is slightly better than DNNBuilder when handling more abundant resources in Scheme 2. However, its folded structure and coarse-grained configuration prevent further scaling in Scheme 3, and it generates an accelerator with the same size as that from Scheme 2.

\section{F-CAD Automation Design Flow}

\begin{figure}[t]
    \centering
    \includegraphics[width=0.49\textwidth]{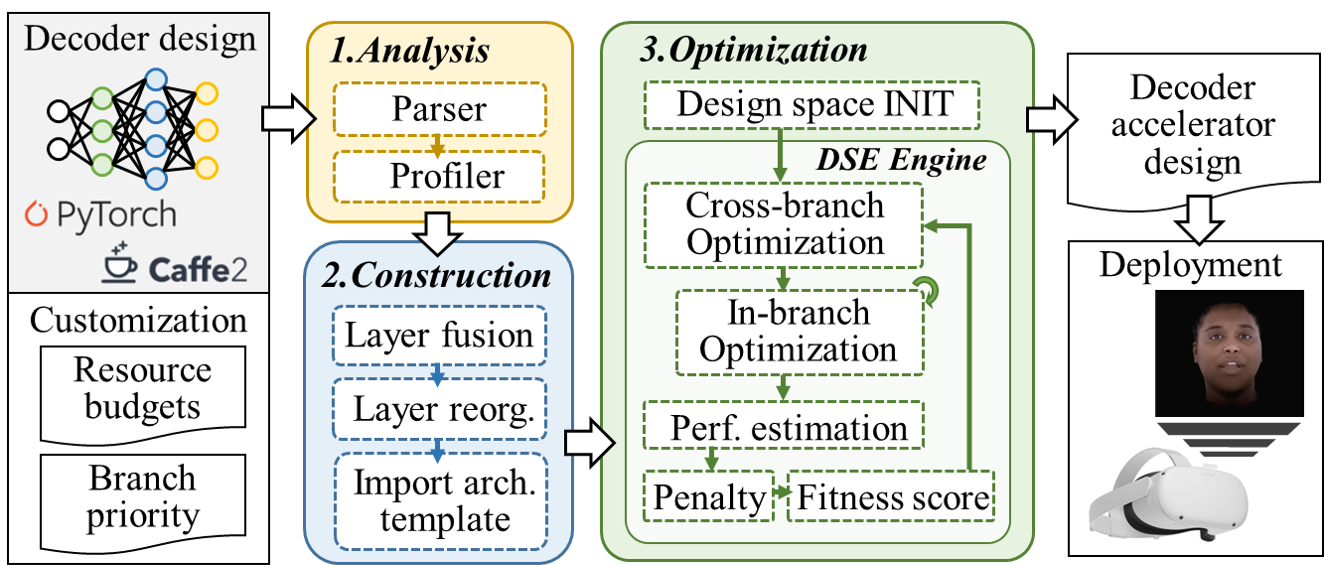}
    \vspace{-10pt}
    \caption{The proposed F-CAD design flow with three major steps 
    to deliver optimized hardware accelerators for codec avatar decoding.}
    \label{fig:designflow}
    \vspace{-5pt}
\end{figure}

To address these challenges, we propose F-CAD to design and explore customized hardware accelerators for multi-branch DNNs and leveraging codec avatar decoding with high throughput and efficiency. As shown in Fig. \ref{fig:designflow}, F-CAD directly connects to popular machine learning frameworks and takes the developed decoder models as inputs as well as arbitrary hardware budgets and differentiated branch priority for more refined customization. 

In \textit{Analysis} step, F-CAD starts analyzing the targeted network by extracting not only layer-wise information (e.g., layer types, layer configurations), but also branch-wise information (e.g., branch number, number of layers in each branch, and layer dependencies). Then, the profiler begins to calculate the compute and memory demands of each layer and provides statistics on branch-wise demands 
to help mapping the targeted decoder onto our proposed accelerator architecture. Inputs also contain resource budgets and branch priority to set up resource boundaries and highlight the importance of different branches for architecture exploration in Step 3.

In \textit{Construction} step, layer fusion is performed to reduce the layer number, where lightweight layers (e.g., activation layers) are aggregated to their neighbouring major layers, such as Conv-like and up-sampling layers, which dominate the computation or memory consumption. 
Branches with shared parts are then separated to create individual dataflows and the corresponding layers are reorganized and assigned to 
the flow with the highest computation demand. This strategy helps avoid hardware redundancy as no duplicated hardware units will be instantiated, and it creates a clear critical flow (with the most computations) from the shared branches and guarantees this flow will get enough attentions in the \textit{Optimization} step.
For example, Br. 2 and 3 of the targeted decoder share the same front-end, layers from this part will be assigned to Br. 2 as it is more critical and contains higher computation demand. After fusion and reorganization, F-CAD imports and expands the proposed elastic architecture (Sec. \ref{sec:arch}) along X and Y dimensions according to the layer and branch number, respectively. Eventually, a basic accelerator is generated and will be optimized in Step 3.

In \textit{Optimization} step, the accelerator design space (Sec. \ref{sec:design_space}) is first determined.
The layers and branches of the decoder contribute to the higher dimensional design space so it becomes complex to search for the optimized design. F-CAD introduces a DSE engine (Sec. \ref{sec:dse}) to leverage both cross-branch and in-branch optimization. A stochastic search is applied in the cross-branch optimization to explore resource distribution schemes across branches; while a greedy search is adopted for in-branch optimization, finding the best accelerator candidate for each branch by considering design spaces and available resources.
After that, accelerator candidates are evaluated against performance, efficiency, and customization requirements. 
The DSE engine eventually generates the globally optimized design through an iterative process.
  
\begin{figure*}[t]
\centering
    \includegraphics[width=0.95\textwidth]{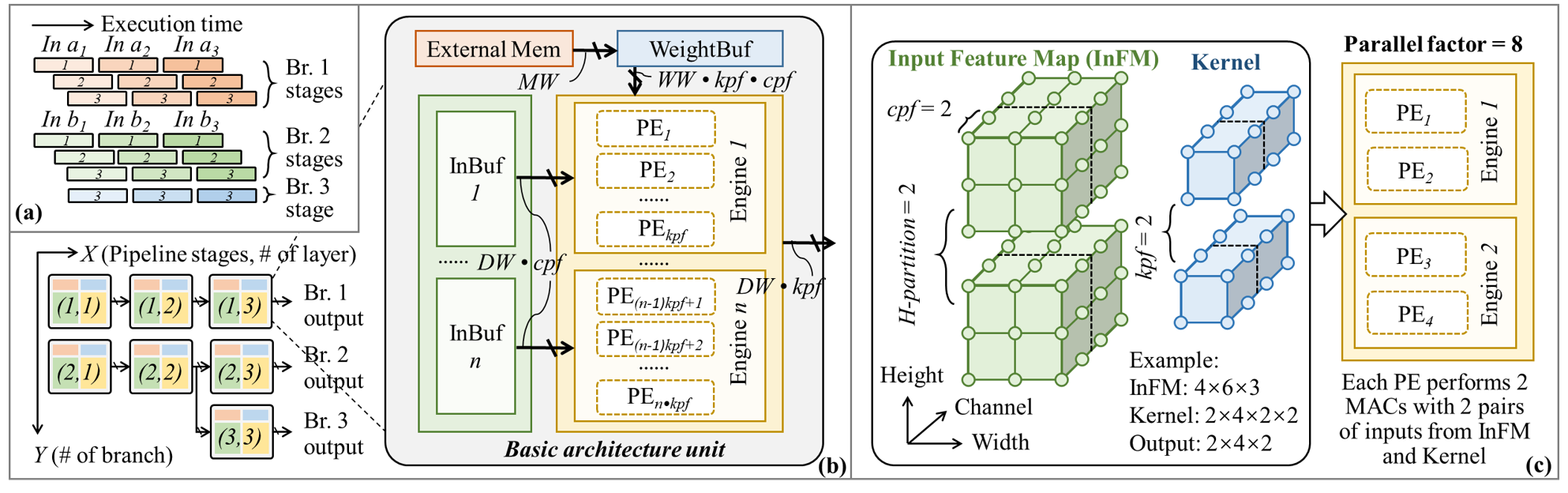}
    \vspace{-5pt}
    \caption{(a) F-CAD generated accelerators follow a layer-based multi-pipeline accelerator paradigm, where each layer after reorganization contributes to a pipeline stage; (b) The proposed elastic architecture features a \textbf{flexible expansion capability}, which can be extended along \textit{X}- and \textit{Y}-axis to instantiate multiple basic architecture units according to the branch number and the layer number in each branch for parallel processing across branches. Each basic architecture unit is equipped with external memory, on-chip memory (for weight buffers and input buffers), and computation resources (for PEs). It provides maximum \textbf{three-dimensional parallelism (3D parallelism)} following input channel, output channel, and the height of the input feature map when handling compute-intensive DNN layers; (c) Example of data partitioning to enable the proposed \textbf{3D parallelism}. }
    \label{fig:arch}
    \vspace{-5pt}
\end{figure*}

\section{Accelerator Architecture}
\label{sec:arch}
\subsection{A layer-based multi-pipeline accelerator paradigm}
The design paradigm of our proposed accelerator is presented in Fig. \ref{fig:arch} (a). Inputs of each branch (e.g., three 256-dimensional latent codes named \textit{In} $a_1$ $\sim$ $a_3$ for branch 1) are processed in a pipeline manner and passed through all pipeline stages belonging to that branch. For branches with the shared part, corresponding stages are assigned to one of the branches following the layer reorganization strategy. 
For example, Br. 2 and 3 share the first two layers, so stage 1 $\sim$ 2 are assigned to Br. 2 in this case, while the subsequent stages are separately executed. The results of stage 2 are distributed to two different branches. We also adopt the fine-grained pipeline design from \cite{zhang2018dnnbuilder} to lower the pipeline initial latency.

\subsection{The elastic architecture with 2D expansion capability}
To enable the proposed accelerator paradigm, F-CAD introduces an elastic architecture to flexibly expand the accelerator following two dimensions. In Fig. \ref{fig:arch} (b), this elastic architecture consists of basic architecture units that are arranged in a two-dimensional plane reflecting the layer reorganization results and each unit is responsible for one pipeline stage. For example, the expansion following the \textit{X}-axis, such as unit instances (1,1), (1,2), and (1,3), means more stages (three in this case) need to be handled in this branch; while the expansion along the \textit{Y}-axis represents more branches are used in the targeted decoder. In this example, F-CAD generates an accelerator with three pipelines corresponding to Br. 1 $\sim$ 3. 

Inside the basic architecture unit, there are three types of resources as: computation (yellow area), on-chip memory (blue and green areas), and external memory (red area) resources. The input feature map from the previous layer is passed horizontally from the left and a fraction of it is kept in the input buffer (InBuf) to provide timely data supply. Meanwhile, DNN parameters 
are fetched from external memory and stored in the weight buffer (WeightBuf) following the computation order. Each basic architecture unit is highly configurable to meet various requirements from different layer stages. It supports the proposed 3D parallelism, 
which includes two unrolling factors along the output and input channels (kernel parallelism factor \textit{kpf} and channel parallelism factor \textit{cpf}) and the partition factor of the input feature map (\textit{H-partition}). After configuration, \textit{H-partition} number of compute engines are instantiated and each engine contains \textit{kpf} process elements (PEs) to handle computations. The proposed basic architecture unit also allows customized bitwidth of the input features (DW), the weights (WW), and the external memory bus (MW).

\subsection{The basic architecture unit with 3D parallelism }
Fig. \ref{fig:arch} (c) provides a detailed illustration of the proposed 3D parallelism. Assuming a Conv-like layer with a 4$\times$6$\times$3 input feature map (InFM) and two 4$\times$2$\times$2 kernels. 
We have the maximum input parallel factor as \textit{cpf}$_{max}=4$ and the maximum output parallel factor as \textit{kpf}$_{max}=2$, since this layer contains four input channels and two output channels that can be processed in parallel. In this case, we configure both input and output parallel factors as 2 (\textit{cpf} $=$ \textit{kpf} $=2$), so each compute engine will instantiate two PEs and each PE performs two multiply-accumulations (MACs) in parallel. Since the parallelism from input/output channels may not be sufficient for codec avatar decoding, we add one additional parallelism by partitioning the InFM along the height dimension. So, all InFM subsections can be processed in parallel. The total parallel factors of this example are \textit{cpf} $\times$ \textit{kpf}$ \times$ \textit{H-partition} $=8$ with four PEs instantiated. 

\section{Multi-branch Design Space Exploration}

\begin{table}[t]
\footnotesize
\caption{The multi-branch dynamic design space}
\vspace{-20pt}
\label{tab:designspace}
\begin{center}
\newcommand{\tabincell}[2]{\begin{tabular}{@{}#1@{}}#2\end{tabular}}
\begin{tabularx}{0.49\textwidth}{p{5pt}|l}
\toprule
\textbf{Br.} & \textbf{Hardware configurable parameters}\\ \hline
1 & $config^1\leftarrow batchsize^1$, $cpf^1_1 \cdots cpf^1_l$, $kpf^1_1 \cdots kpf^1_l$, $h^1_1 \cdots h^1_l$ \\ \hline
2 & $config^2\leftarrow batchsize^2$, $cpf^2_1 \cdots cpf^2_m$, $kpf^2_1 \cdots kpf^2_m$, $h^2_1 \cdots h^2_m$ \\ \hline
$\cdots$ & $\cdots \cdots$\\  \hline
$B$ & $config^B\leftarrow batchsize^B$, $cpf^B_1 \cdots cpf^B_n$, $kpf^B_1 \cdots kpf^B_n$,\\ 
& \quad\quad\quad\quad\quad\quad  $h^B_1 \cdots h^B_n$ \\ \hline\hline
\end{tabularx}

\begin{tabularx}{0.49\textwidth}{c|c}
Customization & $Q$, $BatchSize_1 \cdots BatchSize_B$, $P_1 \cdots P_B$  \\ \hline
Resource budgets & $C_{max}$, $M_{max}$, $BW_{max}$  \\
\bottomrule
\end{tabularx}
\end{center}
\vspace{-15pt}
\end{table}

\subsection{Multi-branch dynamic design space}
\label{sec:design_space}
Although the highly configurable and scalable features of the proposed elastic architecture help address the unique network structures adopted by codec avatar decoders, they also introduce a complicated and high-dimensional design space. 
The more branches in the decoder, or more layers in a branch there are, the 
higher dimensional design space it becomes. 
We define it as a multi-branch dynamic design space and summarize all configurable hardware parameters in Table \ref{tab:designspace}. 
Assuming a decoder with $B$ branches, we pick the first two (with $l$ and $m$ layers) and the last branch (with $n$ layers) as examples. Each of them can be configured in four major aspects as the \textit{cpf}, \textit{kpf}, and \textit{H-partition} ($h$) from the proposed 3D parallelism and batch size ($batchsize$). Parameters of the same branch are passed to a configuration file ($config$) and all 
these files together describe the overall accelerator configuration. 
The goal of F-CAD is to explore the best accelerator configuration in the design space by considering the customization and resource constraints. The customization includes the data quantization ($Q$), the branch-wise targeted batch size ($BatchSize$), and the priority ($P$) to indicate different importance of each branch. While the resource budgets specify three major resources as compute resource $C_{max}$, on-chip memory $M_{max}$, and external memory access bandwidth $BW_{max}$.

\subsection{Design space exploration}
\label{sec:dse}
To effectively search for the best configuration, the proposed DSE engine adopts a two-step strategy with a cross-branch stochastic search and an in-branch greedy search. It follows the divide and conquer idea, to first confirm the resource distribution for every branch and then aim for the best individual branch configuration with given resources.

\subsubsection{\textbf{Cross-branch optimization}}
In Algorithm 1, the proposed DSE engine first randomly generates $\mathcal{P}$ resource distribution schemes ($rd$). Each scheme is considered as a candidate, corresponding to a cross-branch resource distribution regarding compute resource $C$, on-chip memory $M$, and external memory access bandwidth $BW$. In each iteration, $rd$ is then passed to Algorithm 2 for a detailed hardware configuration ($Config$) following the proposed architecture template.
With the $Config$, we can evaluate the accelerator performance and calculate the fitness score for every candidate. We build a function \textbf{S} to provide a weighted score based on the branch performance $Perf=$\{$perf_1 \cdots perf_B$\} and branch priority factor as
$\sum_{j=1}^{B}perf_j \times P_j$. 
We also introduce a penalty term \textbf{P} to control the branch-wise performance variance as: 
$\alpha \times \sigma^2(Perf)$.
Then, we calculate the fitness score by subtracting \textbf{P} from \textbf{S}. A candidate with higher fitness score means it is better than others. We define $rd^{best}_i$ and $rd^{best}_{global}$ to keep track of the local best of each candidate across all iterations and the global best candidates. $rd^{best}_i$ and $rd^{best}_{global}$ can clarify the optimization directions in each iteration, 
so that each candidate can be evolved iteratively to approach the local and the global best positions by a random distance. 
By performing such a stochastic search, 
eventually, Algorithm 1 discovers the global best design by considering the given constraints.

\begin{algorithm}[t!]
\renewcommand\baselinestretch{0.9}\selectfont
\footnotesize
\caption{Cross-branch optimization algorithm}
Setup resource budgets: $budget=$\{$C_{max}$, $M_{max}$, $BW_{max}$\} \\
Setup maximum iteration number: $\mathcal{N}$\\
Import user customization: $\mathcal{U}=$ \{$BatchSize$, $Priority$\}, where $BatchSize=$\{$BatchSize_1, \cdots, BatchSize_B$\}, $Priority=$\{$P_1, \cdots, P_B$\}\\
Randomly initialize $RD^0$ with $\mathcal{P}$ population: \{$rd^0_1, \cdots, rd^0_p$\} \\
\SetInd{0.5em}{0.5em}
\For{$iter$ in range($\mathcal{N}$)}
{
    \For{$rd^{iter}_i$ in $RD^{iter}$}
    {
        \For{$br_j$ in \{$br_1, \cdots, br_B$\}} 
        {
            $config^j \leftarrow$ \textbf{InBranchOptim}($rd^{iter}_i$, $\mathcal{U}$) \Comment*[f]{\textrm{Algorithm 2}}
        }
        $Config =$ \{$config^1, \cdots, config^B$\} \\
        $Perf$ $\leftarrow$ \textbf{Eval}($Config$) \Comment*[f]{\textrm{Evaluate performance}} \\
        $fitness\leftarrow$ \textbf{S}($Perf$, $\mathcal{U}$) $-$ \textbf{P}($Perf$) \Comment*[f]{\textrm{Get fitness score}}\\
        $rd^{best}_i$, $rd^{best}_{global} \leftarrow$ \textbf{Update}($fitness$, $rd^{best}_i$, $rd^{best}_{global}$)\\
        \uIf{$rd^{best}_{global}$ has changed}{
            $Config^{best}_{global} = Congfig$ \Comment*[f]{\textrm{Save the best HW config.}}
        }
        $rd^{iter+1}_i \leftarrow$ \textbf{Evolve}($rd^{iter}_i$, $rd^{best}_i$, $rd^{best}_{global}$, $budget$)
    }
} 
\textbf{return} $rd^{best}_{global}$, $Config^{best}_{global}$ \Comment*[f]{\textrm{Output the global optimal design}}
\label{alg:crossbranch}
\end{algorithm}

\begin{algorithm}[t]
\renewcommand\baselinestretch{0.9}\selectfont
\footnotesize
\caption{In-branch optimization algorithm}
Input resource distribution: $rd=$ \{$C$, $M$, $BW$\}\\
Input user customization from $\mathcal{U}$: $BatchSize$\\ 
Initialize $config$ for a $l$-layer branch: \{$pf_1, \cdots, pf_l$\}  \\
\SetInd{0.5em}{0.5em}
\For{$layer_k$ in \{$layer_1, \cdots, layer_l$ \}}
{
    $op_k\leftarrow$ \textbf{GetOP}($layer_k$)\\
    $norm\_param_k\leftarrow$ \textbf{GetReuse}($layer_k$)
}
$op_{min} = min(op_1, \cdots, op_l)$\\
$norm\_bw=\sum_{k=1}^{l} (op_k/op_{min})\times norm\_param_k \times Freq$ \\ 
\For{$k$ in range($l$)}
{
    $pf_k= \lceil BW/norm\_bw \times (op_k/op_{min}) \rceil$ \Comment*[f]{\textrm{Parallelism targets}} \\ 
}
\While{True}
{
    \For{$layer_k$ in \{$layer_1, \cdots, layer_l$ \}}
    {
        $cpf_k$, $kpf_k$, $h_k \leftarrow$ \textbf{GetPF}($pf_k$, $layer_k$)\\
        $c_k$, $m_k$, $bw_k \leftarrow$ \textbf{Utilization}($cpf_k$, $kpf_k$, $h_k$)\\
    }
    $batchsize = min(C/\sum_{k=1}^{l}c_k, M/\sum_{k=1}^{l}m_k, BW/\sum_{k=1}^{l}bw_k)$\\
    \uIf{$batchsize < BatchSize$}
    {
       \{$pf_1, \cdots pf_k$\}$/2$
    }
    \Else
    {
        $batchsize = BatchSize$ \textbf{break}
    }
}

$config \leftarrow$ $batchsize$, \{$cpf_1 \cdots cpf_k$\}, \{$kpf_1 \cdots kpf_k$\}, \{$h_1 \cdots h_k$\}  \\
\textbf{return} $config$ \Comment*[f]{\textrm{Output HW config.}}
\label{alg:inbranch}
\end{algorithm}



\begin{table*}[t]
\renewcommand{\arraystretch}{0.95}
\footnotesize
\vspace{-4pt}
\caption{F-CAD generated accelerators for codec avatar decoding}
\vspace{-16pt}
\label{tab:rst}
\begin{center}
\newcommand{\tabincell}[2]{\begin{tabular}{@{}#1@{}}#2\end{tabular}}
\begin{tabular}{c||c|c|c|c|c|c|c|c}
\toprule
   & \textbf{Br.} & \textbf{DSP Usage} & \textbf{Total DSPs} & \textbf{BRAM Usage} & \textbf{Total BRAMs} & \textbf{FPS} &  \textbf{Efficiency (\%)} & \textbf{DSE Running Time (s)}\\ \hline
\textbf{Case 1:} Z7045 (8-bit)      & 1 & 199 & \multirow{3}{*}{737 (81.8\%)} & 221 & \multirow{3}{*}{884 (81.1\%)} & 61.0 &  76.6 & \multirow{3}{*}{101.8}\\ \cline{2-3} \cline{5-5} \cline{7-8}
Resource budget: & 2 & 500 & & 551 & & 30.5 & 86.6\\ \cline{2-3} \cline{5-5} \cline{7-8}
900 DSPs, 1090 BRAMs & 3 & 38  & & 112 & & 61.0 & 84.2\\
\hline
\hline
\textbf{Case 2:} ZU17EG (8-bit)      & 1 & 351 & \multirow{3}{*}{1357 (83.5\%)} & 280 & \multirow{3}{*}{1024 (64.3\%)} & \textbf{122.1}  & 86.8 & \multirow{3}{*}{77.3}\\ \cline{2-3} \cline{5-5} \cline{7-8}
Resource budget: & 2 & 936 & & 642 & & 61.0 & 92.6\\ \cline{2-3} \cline{5-5} \cline{7-8}
1590 DSPs, 1592 BRAMs & 3 & 70  & & 102 & & \textbf{122.1} & 91.4\\
\hline

\textbf{Case 3:} ZU17EG (16-bit)      & 1 & 351 & \multirow{3}{*}{1301 (81.8\%)} & 382 & \multirow{3}{*}{1573 (98.8\%)} & 61.0 & 86.8 & \multirow{3}{*}{82.8}\\ \cline{2-3} \cline{5-5} \cline{7-8}
Resource budget: & 2 & 928 & & 983 & & 30.5 & 93.4\\ \cline{2-3} \cline{5-5} \cline{7-8}
1590 DSPs, 1592 BRAMs & 3 & 22  & & 208 & & 15.3 & 72.7\\
\hline
\hline

\textbf{Case 4:} ZU9CG (8-bit)      & 1 & 351 & \multirow{3}{*}{2229 (88.5\%)} & 280 & \multirow{3}{*}{1168 (64.0\%)} & \textbf{122.1} &  86.8 & \multirow{3}{*}{\textbf{56.9}}\\ \cline{2-3} \cline{5-5} \cline{7-8}
Resource budget: & 2 & 1808 & & 786 & & \textbf{122.1} & 95.8\\ \cline{2-3} \cline{5-5} \cline{7-8}
2520 DSPs, 1824 BRAMs & 3 & 70  & & 102 & & \textbf{122.1} & 91.4\\
\hline

\textbf{Case 5:} ZU9CG (16-bit)      & 1 & 351 & \multirow{3}{*}{2213 (87.8\%)} & 382 & \multirow{3}{*}{1735 (96.1\%)} & 61.0 &  86.8 & \multirow{3}{*}{67.6}\\ \cline{2-3} \cline{5-5} \cline{7-8}
Resource budget: & 2 & 1792 & & 1183 & & 61.0 & \textbf{96.7}\\ \cline{2-3} \cline{5-5} \cline{7-8}
2520 DSPs, 1824 BRAMs & 3 & 70  & & 188 & & 61.0 & 91.4\\

\bottomrule
\end{tabular}
\vspace{-12pt}
\end{center}
\end{table*}

\subsubsection{\textbf{In-branch optimization}}
$rd$ is passed to Algorithm 2 for the best in-branch hardware configuration.
Since the proposed accelerator follows an unfolded pipeline architecture, its throughput can be maximized when all pipeline stages are load-balanced with similar latency. To achieve this goal, we capture the layer-wise compute demands ($op$) and data reuse characteristics ($norm\_param$) to obtain the most optimistic layer-wise parallelism targets ($pf$) by exhausting the allocated bandwidth resources. After that, a greedy search algorithm is applied to approach hardware configurations ($config$) with the largest level of parallelism under resource constraints. 
It will converge once the parallelism fails to grow. 

\subsubsection{\textbf{Performance estimation}}
\begin{figure}[t!]
\vspace{-10pt}
    \centering
    \includegraphics[width=0.4\textwidth]{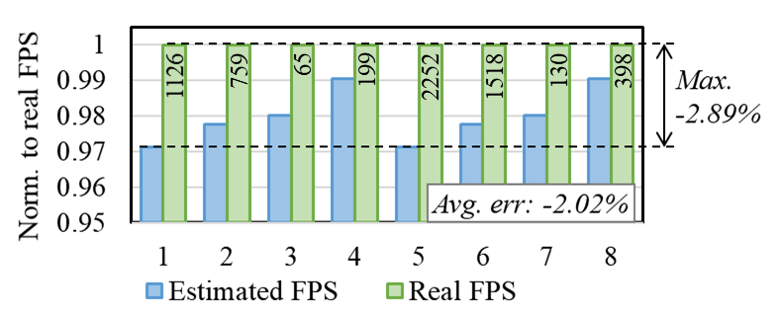}
    \vspace{-10pt}
    \caption{F-CAD FPS estimation errors targeting eight benchmark DNNs.}
    \label{fig:err_fps}
    \vspace{-8pt}
\end{figure}

\begin{figure}[t!]
    \centering
    \includegraphics[width=0.4\textwidth]{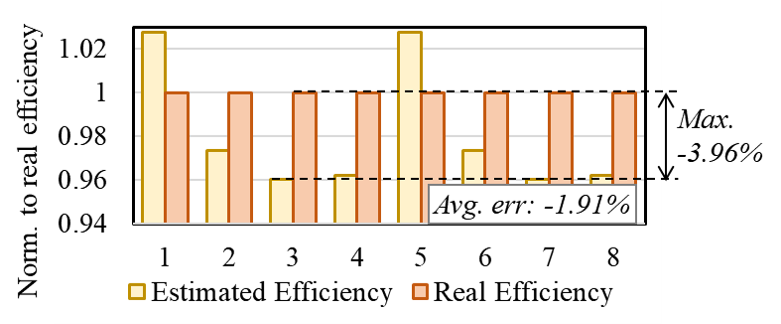}
    \vspace{-8pt}
    \caption{F-CAD efficiency estimation errors targeting eight benchmark DNNs.}
    \label{fig:err_eff}
    \vspace{+6pt}
\end{figure}

We adopt highly accurate analytical models to provide performance and resource utilization feedback and help the DSE engine make the most suitable decisions. Since each branch 
is individually evaluated, we take a branch with $l$ Conv-like layers as an example. For layer $i$, we assume the input feature map size $InCh_i\times H_i \times W_i$ and the kernel size $OutCh_i\times InCh_i\times K_i \times K_i$. With hardware configuration $Config$, the latency $Lat_i$ when executing layer $i$ with working frequency $f$ can be determined as:
\begin{equation}
\small
    Lat_i = \frac{OutCh_i \times InCh_i \times H_i \times W_i \times K_i \times K_i }{cpf_i \times kpf_i \times h_i \times f}
\end{equation}
The overall throughput (FPS) of this branch is:
\begin{equation}
\small
    FPS = \frac{BatchSize}{max(Lat_1, Lat_2, ..., Lat_l)} 
\end{equation}
Estimations also include the resource utilization \{$C$, $M$, $BW$\} (by summing up the resource consumed by all layers) and efficiency (by following Eq. \ref{eq:dsp_eff}). 
To verify the accuracy of our method, we select DNN benchmarks including AlexNet, ZFNet, VGG16, and Tiny-YOLO with 16-bit (benchmarks 1 $\sim$ 4) and 8-bit (benchmarks 5 $\sim$ 8) quantization schemes and compare their estimated performance to the real performance after board-level implementation on a Xilinx KU115 FPGA. As shown in Fig. \ref{fig:err_fps}, we normalize the FPS to the real results in every case to better illustrate the error rate. 
Real FPS results are also listed in the green bars. 
The maximum error is only 2.89\% while the average error is 2.02\%. Similarly, we present the efficiency error in Fig. \ref{fig:err_eff} with 3.96\% maximum error and 1.91\% average error.

\setlength{\textfloatsep}{0pt}

\section{Experimental Results}
In this section, we target three embedded FPGA platforms (Xilinx Z7045, ZU17EG, and ZU9CG) to demonstrate F-CAD's capability and scalability for accelerating codec avatar decoding. Since the targeted platforms are FPGAs, we setup resource budgets $C_{max}$ and $M_{max}$ as the available DSPs and BRAMs in the targeted FPGA, and $BW_{max}$ as the DDR3 memory bandwidth. The clock frequency is set to 200MHz for all platforms. The targeted decoder is described in Table \ref{tab:targeted_decoder} with customized batch size \{$1,2,2$\} corresponding to Br. 1 $\sim$ 3. Such customization is considered by most VR avatar applications where Br. 2 and 3 need to render two HD textures with specular effects seen by both eyes, while the Br. 1 only outputs one facial geometry that can be shared by both eyes. 

Experimental results are listed in Table \ref{tab:rst}, where F-CAD generates five accelerators following the proposed elastic architecture. To evaluate the search speed, we perform 10 independent searches with $\mathcal{N}=20$ (meaning the search contains 20 iterations) and $\mathcal{P}=200$ (meaning 200 resource distribution candidates are initialized) for each case and all of them converge in minutes using an Intel i7 CPU working at 2.6 GHz.
The average iteration for convergence is 9.2 (min: 6.8; max: 13.6).
Eventually, F-CAD generates optimized designs by considering customization and resource constraints. 
In particular, the accelerator for case 4 reaches the highest 122.1 FPS, which fully satisfies the VR requirements; while accelerator for case 5 delivers the highest efficiency peaking at 96.7\%, which can efficiently leverage codec avatar decoding using lightweight HMDs.
%

\begin{table}[t]
\footnotesize
\renewcommand{\arraystretch}{0.95}
\caption{Result comparison to existing accelerators (@200MHz)}
\vspace{-18pt}
\label{tab:rst_comp}
\begin{center}
\newcommand{\tabincell}[2]{\begin{tabular}{@{}#1@{}}#2\end{tabular}}
\begin{tabular}{c|c|c|c|c}
\toprule
    & \textbf{DNNBuilder}\cite{zhang2018dnnbuilder} & \textbf{HybridDNN}\cite{ye2020hybrid} & \multicolumn{2}{c}{\textbf{F-CAD (our work)}}\\ \hline
Precision & 8-bit & 16-bit & 8-bit & 16-bit \\  \hline
DSP & 1820 & 1024 & 2229 & 2213 \\  \hline
BRAM & 1197 & 1120 & 1168 & 1735 \\  \hline
FPS & 30.5 & 22.0 & \textbf{122.1} & \textbf{61.0} \\  \hline
Efficiency & 28.8\% & 70.4\% & \textbf{91.3\%} & \textbf{91.6\%} \\  
\bottomrule
\end{tabular}
\end{center}
\end{table}

We compare F-CAD generated accelerators to existing designs in Table \ref{tab:rst_comp} by targeting the same ZU9CG FPGA with 2520 DSPs and 1824 BRAMs. We use the same mimic decoder mentioned in Sec. \ref{sec:challenges} for DNNBuilder and HybridDNN, while using the targeted decoder (a real-life decoder) for F-CAD. The batch size is uniformly set to one for fair comparison as DNNBuilder and HybridDNN do not support differentiated batch scheme.
The performance and efficiency of DNNBuilder are limited by the insufficient parallelism, so the allocated resources are not fully utilized. On the other hand, HybridDNN fails to allocate more DSPs and leaves more than half of available DSPs unallocated. The reason is that the coarse-grained configuration requires double-sized accelerator instance to continue scaling, but the BRAM budget is not enough and becomes a bottleneck. 
In our design, F-CAD delivers the highest FPS and efficiency given the same resource budgets. Compared to DNNBuilder, we achieve 4.0$\times$ higher throughput and 62.5\% higher efficiency for running the 8-bit codec avatar decoder. Compared to HybridDNN, we can deliver 2.8$\times$ higher throughput by allocating only 2.2$\times$ more DSPs and 21.2\% higher efficiency when running the 16-bit model.
F-CAD can also target ASIC designs with the resource budgets \{$C_{max}$, $M_{max}$, $BW_{max}$\} associating to three most commonly used resources in ASIC DNN accelerators: the available MAC units, the on-chip buffer size, and the external memory bandwidth.

\section{Related Works}
VR telepresence is a recently developed technology that can reproduce authentic human presence including real-time expressions in VR environments to reform remote communication \cite{seymour2017meet,frueh2017headset,lombardi2019neural}. As efforts on improving the telepresence experience, authors in \cite{lombardi2018deep} introduce a DNN-based VAE framework to first encode human's facial geometry and view-dependent appearances into latent codes and then reproduce the corresponding avatar. To enable high quality avatars, authors in \cite{wei2019vr} adopt a generative adversarial network (GAN) in the decoder to create synthetic images while the work proposed in \cite{chu2020expressive} focuses on producing more robust and accurate facial animations. In \cite{richard2020audio}, audio data is also captured for driving realistic facial expressions.
With the compute- and memory-demanding designs, telepresence with codec avatar urgently needs hardware acceleration to guarantee authentic presence with sufficient visual quality when deploying onto hardware-restricted HMDs. 
We have seen recently published works begin to target VR applications from the perspective of hardware and system design, such as \cite{iscaxie2019oo,iscaleng2019energy}. However, the multi-GPU system proposed by \cite{iscaxie2019oo} can not be accommodated by VR headsets, while the VR video processing methods proposed by \cite{iscaleng2019energy} have not addressed the challenges from compute- and memory-intensive DNN workloads (e.g., our targeted VR avatar decoding).
Although DNN hardware accelerators have been designed for accelerating different workloads with diverse hardware devices \cite{qiu2016going,isscc_2016_chen_eyeriss,jouppi2017datacenter,franklin2017nvidia,xu2020autodnnchip,zhang2020dnnexplorer,qin2021}, none of them have dedicated optimization strategies to address the unique challenges of running VAE codec avatar models for VR avatar applications and satisfy their increasing demands. 

\section{Conclusions}
In this paper, we presented F-CAD, an automation tool to design and explore optimized hardware accelerators for VR avatar decoding with high throughput and efficiency. To address the unique challenges coming from the special DNN structure and demanding performance requirements, we proposed an expandable elastic architecture to support multi-branch DNNs and a highly configurable basic architecture unit to provide flexible and scalable parallel processing. We then introduced a multi-branch dynamic design space to describe hardware configurations
and an efficient DSE engine to explore the optimized accelerator by considering various customized constraints and available resource budgets. 
F-CAD delivered the highest throughput and efficiency, peaking at 122.1 FPS and 91.6\%. Compared to the state-of-the-art accelerators, F-CAD achieved 4.0$\times$ and 2.8$\times$ higher throughput and 62.5\% and 21.2\% higher efficiency than DNNBuilder and HybridDNN when targeting the same FPGA.

\bibliographystyle{unsrt}
\bibliography{ref}

\end{document}